\documentclass[aps,preprint,amsfonts]{revtex4}                            
\usepackage[]{bm,epsfig}

\def\bR{\bbox{R}}

\newcommand{\clc}[1]{\multicolumn{1}{c}{#1}}
\newcommand{\be}{\begin{equation}}                                              
\newcommand{\ee}{\end{equation}}

\newcommand{\PRL}[1]{Phys. Rev. Lett.\ {\bf #1}}

\begin{document}                                                                

\title{
Long-range interactions between an atom in its ground S state\\ and an open-shell
linear molecule}

\author{\sc Wojciech Skomorowski and
Robert Moszynski\footnote[1]{Author for correspondence;
e-mail:robert.moszynski@tiger.chem.uw.edu.pl}}

\affiliation{\sl Quantum Chemistry Laboratory, Department of Chemistry, University of Warsaw, Pasteura 1,
02-093 Warsaw, Poland}

\begin{abstract}
Theory of long-range interactions between an atom in its ground S state 
and a linear molecule in a degenerate state with a non-zero projection of the
electronic orbital angular momentum is presented. It is shown how the
long-range coefficients can be related to the first and second-order molecular 
properties. The expressions for the long-range coefficients are written in
terms of all components of the static and dynamic multipole polarizability
tensor, including the nonadiagonal terms connecting states with the opposite
projection of the electronic orbital angular momentum. It is also shown 
that for the interactions of molecules in excited states 
that are connected to the ground state by multipolar transition moments
additional terms in the long-range induction energy appear. All these
theoretical developments are illustrated with the numerical results for
systems of interest for the sympathetic cooling experiments: 
interactions of the ground
state Rb($^2$S) atom with CO($^3\Pi$), OH($^2\Pi$), NH($^1\Delta$),
and CH($^2\Pi$) and of the ground state Li($^2$S) atom with CH($^2\Pi$).
\end{abstract}
\maketitle
\newpage

\section{Introduction}
\label{sec1}
Recent developments in laser cooling and trapping techniques have opened
the possibility of studying collisional dynamics at ultralow temperatures.
Atomic Bose-Einstein condensates \cite{Stringari:99} are of crucial importance
in this respect since investigations of the  collisions between ultracold atoms
in the presence of a weak laser field leads to precision measurements of the
atomic properties and interactions. Such collisions may also lead to the
formation of ultracold molecules that can be used in high-resolution
spectroscopic experiments to study inelastic and reactive processes at
very low temperatures, interatomic interactions at very large distances
including the relativistic and QED effects, or the thermodynamic properties
of the quantum condensates of weakly interacting atoms \cite{Julienne:99}.

Recently, 
experimental techniques based on the buffer gas cooling \cite{Doyle:98}
or Stark deceleration \cite{Meijer:99} produced cold molecules with a
temperature well below 1 K. Optical techniques, based on the laser
cooling of atoms to ultralow temperatures and photoassociation
to create molecules \cite{Julienne:87}, reached temperatures 
of the order of a few $\mu$K or lower. Spectacular achievements were 
reported only very recently with the Bose-Einstein condensation of
homonuclear alkali molecules starting from fermionic atoms 
\cite{Grimm:03,Jin:03,Ketterle:03}.

A major objective for the present day experiments on cold molecules
is to achieve quantum degeneracy for polar molecules. 
Two approaches to this problems are used: indirect methods, in which
molecules are formed from pre-cooled atomic gases \cite{indirect},
and direct methods, in which molecules are cooled from room
temperature. The Stark deceleration and
trapping methods pioneered by  Meijer and collaborators \cite{Bethlem:2003}
are the best developed of the direct methods and provide exciting
possibilities for progress towards quantum degeneracy.

Beam deceleration can achieve temperatures around 10 mK. However,
condensation requires sub-microkelvin temperatures. Finding a
second-stage cooling method to bridge this gap is the biggest
challenge facing the field. The most promising possibility is the
so-called
sympathetic cooling, in which cold molecules are introduced into
an ultracold atomic gas and equilibrate with it. Sympathetic
cooling has already been successfully used to achieve Fermi
degeneracy in $^6$Li \cite{DeMarco:1999} and Bose-Einstein
condensation in $^{41}$K \cite{Modugno:2001}. However, it has not
yet been attempted for molecular systems and there are many
challenges to overcome.
In Berlin, Meijer's group has now developed the capability
to trap ultracold $^{87}$Rb atoms for use in sympathetic cooling
\cite{Adela}.
In London, Tarbutt's group has begun to set up experiments
using an ultracold gas of $^7$Li atoms to cool molecules \cite{Tarbutt:09}. 
Thus far, open shell molecules like CO($^3\Pi$), OH($^2\Pi$), NH($^1\Delta$),
and CH($^2\Pi$) could be decelerated, and are good candidates for sympathetic
cooling. Other simple systems like LiH \cite{sk1}, ND$_3$
\cite{pz1,pz2}, and ions, Yb$^+$ 
and Ba$^+$ \cite{Zipkes2010,Zipkes2010a,Schmid,krych}, 
were also investigated, both exprimentally and theoretically.

Very little is known about collisions between polar molecules and
alkali metal atoms, and results of theoretical studies on them are essential to
guide the experiments and later to interpret the results.
There are two essential ingredients: good potential energy
surfaces to describe the interactions, and good methods for
carrying out low-energy collision calculations.
Hutson and collaborators has pioneered the study of
potential energy surfaces for interactions between polar molecules
and alkali metal atoms \cite{pz1,pz2,Lara:07}. 
However, the theoretical methods available for
calculating the surfaces have some significant inadequacies, which
need to be addressed before quantitative predictions will be
possible.

When dealing with collisions at ultra-low temperatures the accuracy of the potential
in the long range is crucial. Therefore, the methods used in the
calculations of the potential energy surfaces should be size-consistent \cite{Bartlett} in order to 
ensure a proper dissociation of the electronic states, and a proper long-range asymptotics 
of the potential should be imposed. 
The latter task is  highly nontrivial when
a molecule is in an open-shell degenerate state. To our knowledge this problem
has only been addressed by Spelsberg \cite{Spelsberg:99} for the CO+OH system,
and by Nielson {\it et al.} \cite{Nielson:76} and  Bussery-Honvault {\it et al.} 
\cite{Bussery:08,Bussery:09} for an atom interacting with
open-shell molecule. The latter considerations  were limited, however, to the $C_6$ coefficients at most.  
Standard approaches based on the symmetry-adapted perturbation theory within the wave 
function \cite{Jeziorski:94,Moszynski:08} or density functional formalisms \cite{Zuchowski:08}
fail in this case.

In this paper we  report a theoretical study of the long-range interactions
between an atom in its ground S state and a linear molecule in a degenerate state with
the projection of the electronic orbital angular momentum $\Lambda$. The expressions derived in
this work are applied to systems of interest for the sympathetic cooling
experiments, i.e. to interactions of the ground
state Rb($^2$S) atom with CO($^3\Pi$), OH($^2\Pi$), NH($^1\Delta$),
and CH($^2\Pi$), and the Li($^2$S) atom with CH($^2\Pi$). 
The plan of this paper is as follows. In sec. \ref{sec2} we present
the derivation of the long-range asymptotics for the interaction of a ground S state
atom with linear molecule in a degenerate state $\Lambda$. We discuss which polarizability
components of an open-shell species are needed to express the asymptotic interaction energy
and how the expression for the interaction energy  depends on the adopted basis (spherical or Cartesian).
In this section we also
show that for molecules in an excited state that is connected to the ground state
by multipolar transition moments, a new term in the long-range expansion appears.
In sec. \ref{sec3} we present the computational approach adopted in the
this paper, discuss our results, and compare them with
the data from  the
supermolecule calculations. Finally, sec. \ref{sec5}
concludes our paper.

\section{Theory}
\label{sec2}
We consider the interaction of an atom A in the ground S state $\psi_A(\rm S)$ and a linear molecule B
in a state $\psi_B(\Lambda)$, where $\Lambda$ is the projection of the electronic orbital angular
momentum of the molecule on the molecular axis. 
All the quantities relating to the atom and the molecule will be designated by subscripts A and B, respectively.    
Since we are interested in the long-range interactions,
the resulting spin multiplicity of the complex do not play any role in our
further developments, and will be omited. The electronic state of the molecule does
not need to be its ground state.
The Hamiltonian $H$ of the complex AB can be written as:
\be 
H=H_0 + V,
\label{H}
\ee
where $H_0$ is the sum of the Hamiltonians describing isolated monomers A and B,
$H_0=H_{\rm A}+H_{\rm B}$, and $V$ is the intermolecular interaction operator collecting
all Coulombic interactions between electrons and nuclei of the monomer A with the
electrons and nuclei of the monomer B. Assuming that the electron clouds of the
monomers do not overlap, $V$ can be represented by the following multipole expansion 
\cite{Wormer:77,Avoird:80,Heijmen:96}:
\be
V=\sum_{l_A,l_B=0}^\infty C_{l_A,l_B}R^{-l_A-l_B-1}
\sum_{m=-l_A-l_B}^{l_A+l_B}(-1)^m Y_{l_A+l_B}^{-m}(\widehat{R})
\left[\widehat{Q}_{l_A}\otimes \widehat{Q}_{l_B}\right]_m^{l_A+l_B}
\label{Vmult}
\ee
where the constant $C_{l_A,l_B}$ is given by
\be
C_{l_Al_B}=(-1)^{l_B}\left[
\frac{4\pi}{2l_A+2l_B+1}
\right]^{1/2} { 2l_A+2l_B\choose 2l_A}^{1/2},
\label{X}
\ee
$Y_l^m(\widehat{R})$ is the normalized spherical harmonic depending on the
spherical angles $\widehat{R}$ of the vector $\bR$ connecting the centers of
mass of the monomers A and B in a space-fixed coordinate system, 
$\widehat{Q}_{l}^{m}$ denotes the multipole
moment operator  in the space-fixed frame.
We made also use of the coupled product of two spherical tensors:
\be
\left[\widehat{Q}_{l_A}\otimes \widehat{Q}_{l_B}\right]_M^L =
\sum_{m_A,m_B}\langle l_A,m_A;l_B,m_B|L,M\rangle\; \widehat{Q}_{l_A}^{m_A}\; \widehat{Q}_{l_B}^{m_B}.
\label{tp}
\ee
where
$\langle l_A,m_A;l_B,m_B|L,M\rangle$ is the Clebsch-Gordan coefficient.

A state $\psi_B(\Lambda)$ of a linear molecule with $\Lambda\ne 0$ is doubly degenerate,
and so is the state of the complex AB at $R=\infty$, which is just a product $\psi_A(\rm S)\psi_B(\Lambda)$.
For the interaction of a ground state S atom with a molecule, the first-order electrostatic 
energy vanishes identically in the multipole approximation,
so the degeneracy is lifted in the second-order, leading to the splitting of
the two states $\psi_A(\rm S)\psi_B(\pm|\Lambda|)$, into the A$^\prime$ and 
A$^{\prime\prime}$ states of the complex AB at finite $R$.
Thus, to obtain the long-range behavior of the A$^\prime$ and A$^{\prime\prime}$ 
states we have  to
diagonalize the second-order interaction matrix:
\be
\boldsymbol{V}^{(2)}=
\pmatrix{ 
V_{\Lambda,\Lambda}^{(2)}&V_{\Lambda,-\Lambda}^{(2)}\cr
V_{-\Lambda,\Lambda}^{(2)}&V_{-\Lambda,-\Lambda}^{(2)}\cr}.
\label{vmatrix}
\ee

The elements $V_{\Lambda,\Lambda'}^{(2)}$ 
of $\boldsymbol{V}^{(2)}$ are given by the standard expressions of the
polarization theory \cite{Cwiok:92} and can be decomposed into the induction and dispersion parts:
\begin{eqnarray}
V&&_{\Lambda,\Lambda'}^{(2)}
= - \sum_{\{k\}}^\infty{^\prime}\frac{\langle \psi_A({\rm S})\psi_B(\Lambda)|V|\psi_A(\{k\})\psi_B(\pm|\Lambda|)\rangle\langle\psi_A(\{k\})\psi_B(\pm|\Lambda|) |V|\psi_A({\rm S})\psi_B(\Lambda')\rangle}
{\omega_{{\rm S},k}}
\cr &&
 -\sum_{\{k\}}^\infty{^\prime}\sum_{\{n\}}^\infty{^\prime}\frac{\langle \psi_A({\rm S})\psi_B(\Lambda) |V|\psi_A(\{k\})\psi_B(\{n\})\rangle\langle \psi_A(\{k\})\psi_B(\{n\})|V|\psi_A({\rm S})\psi_B(\Lambda') \rangle}
{\omega_{{\rm S},k}+\omega_{\Lambda,n}},
\label{Vll'}
\end{eqnarray}
where $\omega_{{\rm S},k}=E_{\{k\}}-E_{\rm S}$  
is the excitation energy  of the atom from the ground state $\psi_A(\rm S)$ to the
 excited state $\psi_A(\{k\})$ characterized by the set of quantum numbers $\{k\}$, and  $\omega_{\Lambda, n}=E_{\{n\}}-E_{\Lambda}$ is  the excitation energy from the  
 state  $\psi_B(\Lambda)$   to the excited state $\psi_B(\{n\})$ of the molecule with 
the set of quantum numbers denoted by $\{n\}$.
In the above equation the sign prime on the summation symbol means that the ground state of the atom or the
states $\psi_B(\pm|\Lambda|)$ of the molecule are excluded from the summations. 

It can  easily be shown \cite{Nielson:76} that in the case of interaction with an atom in S state
the matrix elements $V_{\Lambda,\Lambda}^{(2)}$ and $V_{-\Lambda,-\Lambda}^{(2)}$
are equal, and the same holds for off-diagonal elements $V_{\Lambda,-\Lambda}^{(2)}$ and $V_{-\Lambda,\Lambda}^{(2)}$. 
The eigenfunctions and the corresponding eigenvalues of the matrix (\ref{vmatrix}) can simply be constructed  on  
the symmetry basis only, as
the proper zero-order wave function of the complex AB should have a defined  behaviour under the
reflection in the plane $\sigma$ containing the three atoms: $\sigma\psi_{AB}^{(+)}=\psi_{AB}^{(+)}$
for A$^\prime$ state and $\sigma\psi_{AB}^{(-)}=(-1)\psi_{AB}^{(-)}$ for  A$^{\prime\prime}$ state.  
In our approach the three atoms lie in the $xz$ plane and the transformation between the space-fixed frame 
and the body-fixed frames is chosen in such a way that it leads to the coincidence of the axis $y$ in all relevant 
frames of reference, see Fig. 1. Thus, the symmetry (A$^\prime$ or A$^{\prime\prime}$)
of the zero-order wave function $\psi_{AB}^{(\pm)}$  
characterizes its behaviour under the reflection in the plane $\sigma_{xz}$. It is determined by the symmetry properties of the wave
functions of the two constituent monomers, i.e. $\psi_A(\rm S)$ and $\psi_B(\Lambda)$, with respect to the  reflection 
in the plane $\sigma_{xz}$ of their body-fixed frames, which happens to be the same.   
The symmetry relations for the monomer electronic wave functions are ($\Lambda \ne 0$) \cite{Roeggen:71,Zare}:
\be
\sigma_{xz}\psi_B(\Lambda)= (-1)^\Lambda \psi_B(-\Lambda),\;\;
\sigma_{xz}\psi_A({\rm S})=(-1)^{p_A} \psi_A(\rm S),
\ee
where $p_A$ defines the spatial parity of the atomic wavefunction in the S state.
Therefore, the properly adapted wave functions read:
\begin{equation}
\psi_{AB}^{(\pm)}= \frac{1}{\sqrt{2}}\left[\psi_A({\rm S})\psi_B(\Lambda)
+(-1)^{\Lambda+p_A+f}\psi_A({\rm S})\psi_B(-\Lambda)\right].
\label{f1}
\end{equation}
and the corresponding second-order energies (i.e. eigenvalues of the matrix $\boldsymbol{V}^{(2)}$) are 
\begin{equation}
E^{(2)}_{(\pm)}=V_{\Lambda,\Lambda}^{(2)}+(-1)^{\Lambda+p_A+f} V_{\Lambda,-\Lambda}^{(2)}
\label{e1}
\end{equation}
with $f=0$ for A$^\prime$  and $f=1$ for  A$^{\prime\prime}$ state.
A standard phase convention of Condon and Shortley \cite{Condon} was adopted to define $\psi_B(\Lambda)$
and $\psi_B(-\Lambda)$.
More details on the symmetry properties of the electronic wave funcions of diatomic molecules 
can be found in Refs. \cite{Roeggen:71,Zare,Pack:70}. Here we only want to conclude  that in the 
case of the systems under study  
the A$^\prime$ state corresponds to the combination of the two terms with the minus sing in Eqs. (\ref{f1}) and (\ref{e1}) 
 for Rb+OH($^2\Pi$), Rb+CO($^3\Pi$) and Li+CH($^2\Pi$)
, while the A$^\prime$ state has the plus sign for Rb+NH($^1\Delta$).
The opposite holds for the A$^{\prime\prime}$ states.

The multipole expansion of $V_{\Lambda,\Lambda'}$ is readily obtained by inserting
the multipole expansion (\ref{Vmult}) of the interaction operator $V$ 
into Eq. (\ref{Vll'}) and collecting terms,
as it was done in Refs. \cite{Wormer:77,Avoird:80,Heijmen:96}. 
Specifically, the derivation is based on the well known transformation
properties of the multipole operators from the space-fixed (with the index $m_X$) to the body-fixed (with the index $k_X$)
frame of each monomer (X=A or B):
\begin{equation}
\widehat{Q}_{l_X}^{m_X}=\sum_{k_X=-l_X}^{l_X} 
D^{(l_X)^\star}_{m_X,k_X}(\widehat{R}_X)\widehat{Q}_{l_X}^{k_X},
\label{sfbf}
\end{equation}
where $D^{(l_X)^\star}_{m_X,k_X}(\widehat{R}_X)$ is the Wigner $D$ matrix, and on the
addition theorems for the $D$ functions and spherical harmonics:\\
\begin{eqnarray}
D^{(l_X)^\star}_{m_X,k_X}(\widehat{R}_X) \cdot D^{(l_X')^\star}_{m_X',k_X'}(\widehat{R}_X)&&=
\sum_{L_X}\sum_{K_X=-L_X}^{L_X}\sum_{M_X=-L_X}^{L_X}
\langle l_X,k_X;l_X',k_X'|L_X,K_X\rangle 
\cr
&&
\times
\langle l_X,m_X;l_X',m_X'|L_X,M_X\rangle
D^{(L_X)^\star}_{M_X,K_X}(\widehat{R}_X),
\label{DD}
\end{eqnarray}
\begin{eqnarray}
Y_{l_A+l_B}^{-m}(\widehat{R})\cdot Y_{l_{A'}+l_{B'}}^{-m'}(\widehat{R})&&=
\sum_{L}\sum_{M=-L}^L
\left[\frac{(2l_A+2lB+1)(2l_{A'}+2l_{B'}+1)}{4\pi(2L+1)}\right]^{1/2}
\cr
&&
\times \langle l_A+l_B,-m;l_{A'}+l_{B'},-m'|L,M\rangle \langle l_A +l_B,0;l_{A}'+l_{B}',0|L,0 \rangle
\cr
&&
\times
Y_{L}^{M}(\widehat{R}).
\label{YY}
\end{eqnarray}

The sets of the Euler angles $\widehat{R}_A$ and $\widehat{R}_B$ describe rotations of  the space-fixed frame
to the appropriate body-fixed frames. For our convinience we choose the $Z$ axis  along the 
intermolecular axis connecting the two centers of mass. Then, the angular factor $Y_{L}^{M}(\widehat{R})$ reduces to:
\begin{equation}
Y_{L}^{M}(\widehat{R})=Y_{L}^{M}(0,\phi)=\left(\frac{2L+1}{4\pi}\right)^{1/2}\delta_{M,0}.
\label{Ybf}
\end{equation}
For an atom in an S state the quantum numbers $L_A$, $M_A$ and $K_A$ are zero, so $D^{(L_A)^\star}_{M_A,K_A}(\widehat{R}_A)=1$. 
For an open-shell linear molecule we have $M_B=0$, and the set of the three Euler angles is  $\widehat{R}_B=(0,\theta,0)$, so
\be
 D^{(L_B)^\star}_{0,K_B}(0,\theta,0)=(-1)^{K_B}\sqrt{\frac{(L_B-K_B)!}{(L_B+K_B)!}}P^{K_B}_{L_B}(\cos \theta)
\ee
where $P^{K_B}_{L_B}(\cos \theta)$ denotes the associated Legendre polynomial.
The set of the Euler angles $\widehat{R}_B=(0,\theta,0)$ for a linear molecule is consistent with the 
foregoing requirement of the coincidence of the axis $y$  in 
all relevant frames of reference, cf. Fig. 1. The other possible set would be $\widehat{R}_B=(3\pi/2,\theta,\pi/2)$ 
which leads to the
coincidence of the axis $x$ as it was adopted in Ref. \cite{Nielson:76}. 

It is useful to express the final equations for the elements of the matrix $\mathbf{V}^{(2)}$ in terms of the
static and dynamic multipole polarizabilities of the  atom and molecule. For an S atom we use the standard definition:

\be
\alpha^{ll'}_{mm'}(\omega)=
2\sum_{\{k\}}{^\prime}
\;\omega_{{\rm S},k}\frac{\langle \psi_A({\rm S})|\widehat{Q}_l^m|\psi_A(\{k\}) \rangle
\langle \psi_A(\{k\}) |\widehat{Q}_{l'}^{m'}|   \psi_A({\rm S}) \rangle}
{\omega_{{\rm S},k}^2-\omega^2},
\label{polll'}
\ee
while for an open-shell linear molecule we introduce extra superscripts $\Lambda$ and $\Lambda'$ to distinguish between
diagonal and off-diagonal components:  

\be
^{\Lambda,\Lambda'}\alpha^{ll'}_{mm'}(\omega)=
2\sum_{\{n\}}{^\prime}
\;\omega_{\Lambda,n}\frac{\langle \psi_B(\Lambda)|\widehat{Q}_l^m|\psi_B(\{n\})\rangle
\langle \psi_B(\{n\})|\widehat{Q}_{l'}^{m'}|\psi_B(\Lambda')\rangle}
{\omega_{\Lambda,n}^2-\omega^2}.
\label{polar1}
\ee

The corresponding irreducible polarizabilities are obtained by Clebsch-Gordan coupling:

\be 
^{(\Lambda,\Lambda')}\alpha^{(l l')L}_{K}(\omega)=
\sum_{m,m'}\langle l,m;l',m'|L,K\rangle
^{(\Lambda,\Lambda')}\alpha^{l l'}_{mm'}(\omega).
\label{irredpol}
\ee
The only nonvanishing components of the irreducible polarizability for an atom in the S state are $\alpha^{(l l)0}_0$, while for 
an open-shell linear molecule the nonzero components are $^{\Lambda,\Lambda'}\alpha^{(l l')L}_{K}$ with $K=0$ for $\Lambda=\Lambda'$
and with $K=2\Lambda$ if $\Lambda'=-\Lambda$.

We are now ready to give final expressions for the long-range coefficients expressed in terms of 
the irreducible components of the polarizabilities:
\be
V_{\Lambda,\Lambda'}^{(2)}=-\sum_{l_A,l_B,l_B'}\sum_{L,K}
\xi^{L,K}_{l_A l_B l_B'} R^{-(2+2l_A+l_B+l_B')}\left[^{\Lambda,\Lambda'}C_{l_A l_B l_B'}^{L,K}({\rm ind})+\;{^{\Lambda,\Lambda'}C_{l_A l_B l_B'}^{L,K}}({\rm disp})\right]
P_{L}^{|K|}(\cos \theta),
\label{v2fin}
\ee
where the constant $\xi^{L,K }_{l_Al_Bl_B'}$ is given by:
\begin{eqnarray}
\xi_{l_Al_Bl_{B}'}^{L,K}&&=\left[\frac{(2l_A+2l_B+1)!(2l_A+2l_{B'}+1)!}{(2l_A)!(2l_B)!(2l_A)!(2l_{B'})!}\right]^{1/2}
\pmatrix{ l_A+l_B&l_A+l_{B}'&L\cr 0&0&0\cr }
\cr
&&
\times\left(\frac{2L+1}{2l_A+1}\right)^{1/2}
\sqrt{\frac{(L-K)!}{(L+K)!}}\left\{\begin{array}{ccc} l_B & l_A+l_B & l_A \\
                          l_{A}+l_{B}'  & l_{B}'   & L   \end{array}\right\},
\label{xi}
\end{eqnarray}
and the expressions  in the round and curly brackets are the $3j$ and $6j$ coefficients,
respectively. 
Combining all terms with the same power $n=2l_A+l_B+l_{B}'+2$ in the above expansion, we will get standard long-range coefficients $C_n^{L,K}$:
\be
C_n^{L,K}=\sum_{l_A,l_B,l_B' }^{2l_A+l_B+l_{B}'+2=n} \xi^{L,K}_{l_A l_B l_B'}\left[ {^{\Lambda,\Lambda'}}C_{l_A l_B l_B'}^{L,K}({\rm ind})+\;{^{\Lambda,\Lambda'}}C_{l_A l_B l_B'}^{L,K}({\rm disp})\right],
\ee 
by means of which the asymptotic expansion of Eq. (\ref{e1}) simply reads:
\be
E^{(2)}_{(\pm)}=-\sum_{n=6}^\infty\sum_L \left[\frac{C_n^{L,0}}{R^n}P_{L}^{0}(\cos \theta)+ (-1)^{\Lambda+p_A+f}
\frac{C_n^{L,2\Lambda}}{R^n}P_{L}^{|2\Lambda|}(\cos \theta)\right].
\ee

The dipersion part $^{\Lambda,\Lambda'}C_{l_A l_B l_B'}^{L,K}({\rm disp})$  is
proportional to the Casimir-Polder integral over the  atomic and molecular polarizabilities calculated at
imaginary frequencies:
\be
^{\Lambda,\Lambda'}C_{l_A l_B l_B'}^{L,K}({\rm disp})= \frac{1}{2\pi}\int_0^\infty
\alpha^{(l_Al_A)0}_0(i\omega)\;{^{\Lambda,\Lambda'}\alpha^{(l_Bl_B')L}_{K}}(i\omega){\rm d}\omega,
\label{disp}
\ee  
while the induction term $^{\Lambda,\Lambda'}C_{l_A l_B l_B'}^{L,K}({\rm ind})$ is the product of the 
static polarizability of the atom and permanent multipole moments of the open-shell molecule:

\be
^{\Lambda,\Lambda'}C_{l_A l_B l_B'}^{L,K}({\rm ind})= \alpha^{(l_Al_A)0}_0(0)\left[
\langle \psi_B(\Lambda)|\widehat{Q}_{l_B}|\psi_B(\pm|\Lambda|)\rangle\otimes
\langle \psi_B(\pm|\Lambda|) |\widehat{Q}_{l_B'}| \psi_B(\Lambda')\rangle\right]_{K}^{L}.
\label{ind}
\ee
%In the last equation we made use of the coupled product of two spherical tensors:

%\be
%\left[\widehat{Q}_l\otimes \widehat{Q}_{l'}\right]_K^L =
%\sum_{k,k'}\langle l,k;l',k'|L,K\rangle\; \widehat{Q}_l^k\; \widehat{Q}_{l'}^{k'}.
%\label{tp}
%\ee
%Due to the relation $Y_{l}^{m^\star}(\theta,\phi)=(-1)^m Y_{l}^{-m}(\theta,\phi)$ 
%and the fact that $K$ has to be even and equal to $2\Lambda$, we were able to
%avoid dealing with Legendre polynomials $P_{L}^{K}(\cos \theta)$ with $K<0$ in the final expression, Eq. (\ref{v2fin}).  

A few comments are needed here. The expression for the diagonal term $V^{(2)}_{\Lambda,\Lambda}$
is the same as for the interaction between atom and  linear molecule in a spatially nondegenerate state ($\Sigma$).
The additional term emerging when $\Lambda\ne0$ is the off-diagonal $V^{(2)}_{\Lambda,-\Lambda}$.
It has both induction and dispersion parts. As it was  already stated, 
the only nonvanishing term in $V^{(2)}_{\Lambda,-\Lambda}$ appears for 
$K=2\Lambda$. The lowest order long-range coefficient, to which 
the off-diagonal term contributes to, is $C_n^{L,K}$ with $n=4+2|\Lambda|$ for the dispersion part and
$n=5+2|\Lambda|$ for the induction part. The source of the induction energy in $V^{(2)}_{\Lambda,-\Lambda}$ 
comes from the fact that open-shell linear molecules have an additional independent component of the 
permanent multipole moments,
$\langle\psi(\Lambda)|\widehat{Q}^{2\Lambda}_l|\psi(-\Lambda)\rangle$, in contrast to the $\Sigma$ state molecules 
with only one independent component, namely $\langle\psi(\Lambda)|\widehat{Q}^0_l|\psi(\Lambda)\rangle$.
It is obvious that the second  independent component appears for $l\ge 2|\Lambda|$, therefore in the case of 
the $\Pi$ states there is no induction contribution to the off-diagonal $C_6^{2,2}$ coefficient.
The dispersion terms in $V^{(2)}_{\Lambda,-\Lambda}$ results from the presence of additional  
components of the polarizability terms for the open-shell linear molecules. 
 
In general, the number of independent diagonal terms
 $^{\Lambda,\Lambda}\alpha^{ll'}_{kk'}$ is equal to $2l_{<} +1$ (where $l_{<}$ is the smaller of
$l$ and $l'$).
Each of the non-redundant components $^{\Lambda,\Lambda}\alpha^{ll'}_{kk'}$ comes from the
transitions in the sum  (\ref{polar1}) through the intermediate states $\psi(\{n\})$ with 
the projection of the eletronic angular momentum ranging $|\Lambda|-(2l_< +1)$
to $|\Lambda|+(2l_< +1)$.
Due to the transformation properties,
the number of independent irreducible components $^{\Lambda,\Lambda}\alpha^{(ll')L}_{K}$ 
will also be $2l_< +1$. 

In addition, for open-shell linear molecules, there are nondiagonal 
components of the polarizability tensor $^{\Lambda,-\Lambda}\alpha^{ll'}_{kk'}$
  which do not vanish, and are not related to the diagonal terms. The nondiagonal
terms in the multipole polarizability tensor appear if the condition $k'+k=2\Lambda$
can be fulfilled for a given set of the quantum numbers $(l,k)$ and $(l',k')$. They  result
 from the fact that there are transitions through the same operator to  states $\Sigma^+$
and $\Sigma^-$, and then  $^{\Lambda,-\Lambda}\alpha^{ll'}_{kk}$ is the diffrence between these two separate
contributions (e.g. $^{1,1}\alpha^{11}_{11}$ for the $\Pi$ state). 
The other mechanism leading to the appearance of the  nondiagonal $^{\Lambda,-\Lambda}\alpha^{ll'}_{kk'}$ terms
is parallel with the source of the off-diagonal induction terms, namely there are possible 
two independent transition
moments between the state in question $|\psi(\Lambda)\rangle$ and the exited states through an 
operator of the same rank, 
e.g. $^{1,-1}\alpha^{12}_{02}$  for the $\Pi$ state, to be compared with the diagonal $^{1,1}\alpha^{12}_{00}$ .

Let us discuss  the dipole polarizabilities in more details. For open-shell diatomic molecules there are
three independent spherical components $^{\Lambda,\Lambda}\alpha^{11}_{kk'}$
with $(k,k')$ equal to $(0,0)$, $(1,-1)$ and $(-1,1)$, the corresponding transitions in sum-over-states 
occur through the  excited states with a projection of electronic angular momentum 
equal to   $\Lambda$, $\Lambda-1$, and $\Lambda+1$, respectively. 
In case of the $\Pi$ states there is extra nondiagonal component $^{1,-1}\alpha^{11}_{11}$
with intermediate
states $\Sigma^+$ and $\Sigma^-$ in the sum-over-state expression, which come with opposite sign, 
in contrast to $^{1,1}\alpha^{11}_{-11}$
 in which  states
$\Sigma^+$ and $\Sigma^-$ contribute with the same sign.
The four corresponding irreducible dipole polarizability components would be $^{\Lambda,\Lambda}\alpha^{(11)0}_{0}$,
$^{\Lambda,\Lambda}\alpha^{(11)2}_{0}$, $^{\Lambda,\Lambda}\alpha^{(11)1}_{0}$, and 
$^{\Lambda,-\Lambda}\alpha^{(11)2}_{2}$, the last one is nonvanishing only for the $\Pi$ states.

From the analysis of Eq. (\ref{v2fin}) it turns out that not all nonvanishing irreducible 
components $^{\Lambda,\Lambda'}\alpha^{(ll')L}_{K}$ of the molecule polarizability tensor are needed  
to express the interaction energy of an open-shell linear molecule with an atom. 
The dipole component $^{\Lambda,\Lambda}\alpha^{(11)1}_{0}$
is redundant in this case.     
It follows from the expression (\ref{xi})  that all
terms $^{\Lambda,\Lambda'}\alpha^{(ll')L}_{K}$ with $l+l'+L$  odd  
do not contribute to the interaction energy in the second order $E_{(\pm)}^{(2)}$. 
However, if we had an asymetric molecule or an other
linear open-shell species instead of an atom
then all non-zero irreducible
componets would be  necessary.      
For a $\Delta$ state there are three independent dipole polarizabilities, but again, in the case of 
interaction  with an atom, $^{2,2}\alpha^{(11)1}_{0}$ is redundant.
The first off-diagonal term for a $\Delta$ state will be $^{2,-2}\alpha^{(22)4}_{4}$
or equivalently  $^{2,-2}\alpha^{22}_{22}$.

Sometimes it is more convenient to use the Cartesian basis both for the states $|\Lambda\rangle$ and multipole
moments. The transformation formulas between the two bases can be found
in Ref. \cite{Gray:75}. We focus again on the $\Pi$ states. In the Cartesian  basis the two degenerate 
states $|\pm 1\rangle$ are usually referred to
as $|\Pi_{\rm{x}}\rangle$ and $|\Pi_{\rm{y}}\rangle$.  
The four independent dipole polarizability tensor components would be  $^{\rm{x,x}}\alpha_{\rm{xx}}$,
$^{\rm{x,x}}\alpha_{\rm{yy}}$, $^{\rm{x,x}}\alpha_{\rm{zz}}$,  and $^{\rm{x,y}}\alpha_{\rm{xy}}$,
where we adopted the index (x,y) in place of (-1,1) to distinguish between particular 
diagonal and off-diagonal components. It is 
possible to define the fifth Cartesian component, namely $^{\rm{x,y}}\alpha_{\rm{yx}}$,
however, due to the relation: $^{\rm{x,x}}\alpha_{\rm{xx}}- {^{\rm{x,x}}}\alpha_{\rm{yy}}
= {^{\rm{x,y}}}\alpha_{\rm{xy}}+ {^{\rm{x,y}}}\alpha_{\rm{yx}}$, it will not be  independent \cite{Spelsberg:99}.
To see better why the Cartesian basis may be useful let us mention how the irreducible
spherical components are related to the  Cartesian ones for the $\Pi$ states:
\begin{eqnarray}
^{1,1}\alpha^{(1,1)0}_0&&=^{-1,-1}\alpha^{(1,1)0}_0= -\frac{1}{\sqrt{3}}\left({^{\rm{x,x}}}\alpha_{\rm{xx}}+{^{\rm{x,x}}}\alpha_{\rm{yy}}+{^{\rm{x,x}}}\alpha_{\rm{zz}}\right)\cr
^{1,1}\alpha^{(1,1)2}_0&&=^{-1,-1}\alpha^{(1,1)2}_0=\frac{1}{\sqrt{6}}\left(2\; {^{\rm{x,x}}}\alpha_{\rm{zz}}-{^{\rm{x,x}}}\alpha_{\rm{xx}}-{^{\rm{x,x}}}\alpha_{\rm{yy}}\right)\cr
^{1,1}\alpha^{(1,1)1}_0&&=-^{-1,-1}\alpha^{(1,1)1}_0= -\frac{1}{\sqrt{2}}\left( {^{\rm{x,y}}}\alpha_{\rm{xy}}- {^{\rm{x,y}}}\alpha_{\rm{yx}}\right) \cr
^{1,-1}\alpha^{(1,1)2}_2&&=^{-1,1}\alpha^{(1,1)2}_{-2}= {^{\rm{x,x}}}\alpha_{\rm{yy}}- {^{\rm{x,x}}}\alpha_{\rm{xx}}
\end{eqnarray}
An inspection of the above relations shows that in order to 
express the interaction energy of a $\Pi$ molecule with an atom, only diagonal Cartesian components are needed, as 
the off-diagonal ${^{\rm{x,y}}}\alpha_{\rm{xy}}$ contributes only to  $^{1,1}\alpha^{(1,1)1}_0$, i.e. the term 
not present in expression for $E^{(2)}_{(\pm)}$. 
The off-diagonal irreducible component  $^{1,-1}\alpha^{(1,1)2}_2$ is equal to the difference between two perpendicular
$\alpha$'s calculated for one of the Cartesian states. 
This observation is more general and it turns out that if we decide for a Cartesian representation of the
$|\Lambda\rangle$ states, then all off-diagonal irreducible components $^{\Lambda,-\Lambda}\alpha^{(ll')L}_{2\Lambda}$
 can be related to some diagonal 
Cartesian terms. Still, we will have to calculate some additional Cartesian components, however, diagonal only, 
which are not present in the expression for the diagonal $^{\Lambda,\Lambda}\alpha^{(ll')L}_{0}$ terms. 
Calculations of the off-diagonal polarizability components in the Cartesian basis are indispensable if the interacting 
system comprises of an open-shell molecule and asymmetric species. It means that in such a case the interaction 
energy in the multipole approximation is expressed in terms of, for instance, ${^{\rm{x,y}}}\alpha_{\rm{xy}}$ 
component. It is worth noting that  ${^{\rm{x,y}}}\alpha_{\rm{xy}}$ is irrelevant in the  describtion of the interaction
of the open-shell molecule with the extrenal electric field in the Stark effect. Thus, the induction energy and, consequently, 
intermolecular  forces explores properties of the open-shell diatomic molecules
which are not accessible otherwise.

Note that Eq. (\ref{disp}) is strictly valid only when the molecule
is in its ground electronic state. If the molecule is in an excited state 
that is connected to the ground state (or to any other state lower in energy) by multipolar transition moments
then the Casimir-Polder integral is no longer valid and an extra term has to be added to the energy.   
The reason behind this is the property of the Casimir-Polder integral that if the two
elements in  denominator, $\epsilon_A$ and $\epsilon_B$, have
opposite signs then we obtain (assuming that $\epsilon_B<0$):
\be
\frac{2}{\pi}\int_0^\infty\frac{\epsilon_A}{\epsilon_A^2+\omega^2}\cdot
\frac{\epsilon_B}{\epsilon_B^2+\omega^2}{\rm d}\omega=-\frac{1}{\epsilon_A+|\epsilon_B|},
\ee
instead of the value of $1/(\epsilon_A-|\epsilon_B|)$, which we want to decompose into the product
of two terms depending on monomer properties only.  
Formally, we may write the following identity (again $\epsilon_B<0$):
\be
\frac{1}{\epsilon_A-|\epsilon_B|}=\frac{2}{\pi}\int_0^\infty\frac{\epsilon_A}{\epsilon_A^2+\omega^2}
\cdot\frac{\epsilon_B}{\epsilon_B^2+\omega^2}{\rm d}\omega
+\frac{1}{\epsilon_A+|\epsilon_B|}+\frac{1}{\epsilon_A-|\epsilon_B|}.
\ee
For any molecular state with $\epsilon_B<0$ we have the  summation (\ref{Vll'}) over all atomic states with
positive excitation energy $\epsilon_A$, this means that 
the two last factors in the above equation will add up to yield dynamic polarizability of atom at frequency $\omega=|\epsilon_B|$.
Therefore, if we want to express the whole interaction energy in the second order $E^{(2)}_{(\pm)}$
in terms of the monomers properties only,
then we are forced to add an extra term to the  dispersion part, Eq.  (\ref{disp}), in order to 
compensate an error introduced by
the integral representation of the dispersion energy. 
This additional term 
depends on the dynamic polarizabilites of the atom calculated at frequency $\omega$ equal to the energy of all
possible deexcitations in the molecule, and  will be denoted by $C_{l_A l_B l_B'}^{L,K}({\rm corr,deexc})$.
Its form is slightly  similar to the induction part:

\begin{eqnarray}
^{\Lambda,\Lambda'}C_{l_A l_B l_B'}^{L,K}({\rm corr,deexc})&&= \sum_{\{n^-\}}\alpha^{(l_Al_A)0}_0(\omega_{n^-,\Lambda})
\cr
&&\times\left[
\langle \psi_B(\Lambda)|\widehat{Q}_{l_B}|\psi_B(\{n^-\})\rangle\otimes
\langle \psi_B(\{n^-\}) |\widehat{Q}_{l_B'}| \psi_B(\Lambda')\rangle\right]_{K}^{L}.
\label{inddeexc}
\end{eqnarray}
The summation in the above equation runs only over states of the molecule $\psi_B(\{n^-\})$  
with energy lower than the reference one, i.e. if  $\omega_{n^-,\Lambda}=
E_{\Lambda}-E_{\{n^-\}}$ is positive, and hence $\omega_{n^-,\Lambda}$
corresponds to the possible deexcitations of the molecule.
This term does not have a simple physical interpretation, but as shown in Ref.
\cite{skomo} it leads to a different QED retardation of the long-range
potential than given by the classical Casimir-Polder formula \cite{Polder:48}.
Note also that without this extra term, the second-order interaction energy in the long-range
could not be written correctly in terms of molecular properties of the isolated subsystems
in the case when deexcitation may occur. Obviously, a similar term will be needed if an atom is in an exited state
and molecule in its ground state. Then $ C_{l_A l_B l_B'}^{L,K}({\rm corr,deexc})$ would depend on the
the dynamic polarizabilities of the molecule at frequency $\omega$ corresponding to possible atomic deexcitation.
However, this holds only for atomic excited S states, as if the atomic state was P, D etc. then the whole 
formalism presented here would not be longer valid due to nonvanishing first-order energy in 
the multipole approximation.   

\section{Numerical results and discussion}
\label{sec3}
We have applied the theory exposed above to the interactions of the ground
state rubidium atom Rb($^2$S) with CO($^3\Pi$), OH($^2\Pi$), NH($^1\Delta$),
and CH($^2\Pi$), and of the ground state lithium atom Li($^2$S) with
CH($^2\Pi$).
As discussed in the Introduction, these molecules 
have been successfully decelerated and 
are the best candidates for sympathetic cooling
by collisions with ultracold rubidium atoms. At present no {\em ab initio}
code allows for the calculations of all components of the dynamic polarizability
tensor for open-shell linear molecules. Therefore, in our calculations we
computed the polarizabilities appearing in Eqs. (\ref{ind}) and  (\ref{disp}),
 from the sum-over-states expansion, Eq. (\ref{polar1}). The appropriate 
transitions moments to the excited states and excitation energies were
calculated using linear response formalism with reference wave function obtained
from  the multireference selfconsistent field
method (MCSCF) with large active spaces. For some electronic states the convergence of
the sum in Eq. (\ref{polar1}) was not very fast, and we had to include 
over 100 states in the expansion. The {\sc dalton} program was used for linear response
calculations. We have checked the convergence of the expansion (\ref{polar1}) by
comparison of the static parallel components obtained from the MCSCF
calculations with the results of finite-field restricted open-shell
coupled cluster calculations
with single, double, and noniterative triple excitations, RCCSD(T). The
finite field calculations were done with the {\sc molpro} code \cite{molpro}. 
Note parenthetically that finite field calculations can correctly reproduce
the parallel component of the polarizability tensor, but fail for the
perpendicular component due to the symmetry breaking. The nondiagonal components
cannot be obtained from finite-field calculations.
We have attempted to use the multiconfiguration interaction method restricted to
single and double excitations (MRCI), but due to the convergence problems, 
we could get in this way only a few (up to ten) excited states.
The Li and Rb atoms polarizabilities
at imaginary frequency was taken from highly accurate relativistic calculations
from the group of Derevianko \cite{Derevianko:10}.

In order to judge the quality of the computed long-range coefficients we have
computed cuts through the potential energy surfaces of  
Rb--CH($^2\Pi$) at a fixed distance $R$=30 bohr
from the atom to the center of mass of the molecule. 
The zero of the angle $\theta$ corresponds to the rubidium atom
on the H side of CH.
In these calculations we have employed the supermolecule method.
The potential was computed as the difference,
\be
V^{\rm ^{2S+1}|\Lambda|}(R,\theta)=
E_{\rm AB}^{\rm SM} -
E_{\rm A}^{\rm SM}-E_{\rm B}^{\rm SM},
\label{cccv}
\ee
where $E_{\rm AB}^{\rm SM}$ denotes
the energy of the dimer computed using the supermolecule method SM, and $E_{\rm X}^{\rm SM}$,
X=A or B, is the energy of the atom X.
For the high-spin states (triplet for Rb($^2$S)--OH($^2\Pi$), 
Rb($^2$S)--CH($^2\Pi$), and Li($^2$S)--CH($^2\Pi$), quartet for
Rb($^2$S)--CO($^3\Pi$), and doublet for Rb($^2$S)--NH($^1\Delta$)
we used the restricted open-shell coupled cluster method restricted to
single, double, and noniterative triple excitations 
[RCCSD(T)]. Since the low-spin states have the same asymptotic behavior 
as the high-spin states there was no need
to compute the {\em ab initio} points explicitly.
The RCCSD(T) calculations were done with the {\sc molpro}
suite of codes \cite{molpro}.
The distances in the diatomic molecules were fixed at their equilibrium values $r_e$ corresponding
to the electronic state considered. For CO($^3\Pi$) $r_e$ was set equal to 2.279  bohr, for 
OH($^2\Pi$) 1.834  bohr, for NH($^1\Delta$) 1.954  bohr, and 2.116  bohr for CH($^2\Pi$) \cite{nistpage}.
The angle $\theta=0^\circ$ corresponds to the linear geometris CH--Rb, CH--Li, OH--Rb, NH--Rb and CO--Rb. 
In order
to mimic the scalar relativistic effects some electrons were described
by pseudopotentials. For rubidium we took the ECP28MDF 
pseudopotential from the Stuttgart library \cite{Stoll:05}, and
the $spdfg$ quality basis set
suggested in Ref. \cite{Stoll:05}. 
For the light atoms (hydrogen, carbon, nitrogen, and oxygen)
we used the aug-cc-pVQZ bases \cite{Dunning:94}.
The full basis of the dimer was used in the supermolecule
calculations and the Boys and Bernardi scheme was used
to correct for the basis-set superposition error \cite{Boys:71}.

Before going on with the discussion of the long-range interactions in the dimers, let us compare
the diagonal static polarizabilities of CO($^3\Pi$), OH($^2\Pi$), NH($^1\Delta$),
and CH($^2\Pi$) with the literature data. In fact, the data are very scarce. For OH the most
recent calculations of Spelsberg \cite{Spelsberg:99} date back to 1999 (see also some older
references \cite{Adamowicz:88,Dinur:94,Karna:96}). For CH the only calculation
we found in the literature is the 2007 paper by Manohar and Pal \cite{Pal:07}. To our
knowledge no data for the excited states of CO and NH were reported thus far.
An ispection of Table \ref{tab0} shows a relatively good agreement with the
results of Spelsberg \cite{Spelsberg:99} for OH. The differences are of the order
of a few percent, 5.5\% for the parallel component, and 6.3\% and 3.3\% for the
perpendicular $xx$ and $yy$ components, respectively. For CH Manohar and Pal \cite{Pal:07}
reported only the value of the parallel component obtained from the analytical second derivative
calculations with the Fock space multireference coupled cluster theory restricted to single and
double excitations. The agreement of our result with the value of Ref. \cite{Pal:07}
is remarkably good: the two results agree within 0.4\%.

The long-range coefficients $C_n^{L,K}$ for the interactions of the ground
state rubidium atom Rb($^2$S) with CO($^3\Pi$), OH($^2\Pi$), NH($^1\Delta$),
and CH($^2\Pi$), and of the ground
state lithium atom Li($^2$S) with CH($^2\Pi$) are reported in Tables
\ref{tab1}--\ref{tab5}. Also reported in these tables are the values
of the induction and dispersion coefficients
$C_n^{L,K}({\rm ind})$ and $C_n^{L,K}({\rm disp})$.
First we note that for all systems and most of the coefficients the induction
part is as important as the dispersion. This is not very surprising since
the Li and Rb atoms are highly polarizable, and the molecules suitable
for the Stark deceleration have large
dipole moments. Note parenthetically that the coefficient $C_6^{22}({\rm ind})$ 
for the interactions of the $\Pi$ state molecules and $C_8^{24}({\rm ind})$
for the interactions of the $\Delta$ state molecules vanish for symmetry
reasons.
The leading contribution to the anisotropy of the potentials
in the long range,
as measured by the ratio $C_6^{20}/C_6^{00}$,
is quite substantial for all the systems. The ratio $C_6^{20}/C_6^{00}$
ranges between 0.26 for Li--CH and Rb--CH to 0.43 for Rb--OH. The difference
in the anisotropy due to the presence of terms $C_n^{L 2}/R^n$
is relatively modest, since the coefficients $C_n^{L 2}$ are at
least one order of magnitude smaller than $C_n^{L 0}$.

Comparison of the
long-range anisotropy of the potential energy surfaces for the singlet and triplet
A$^\prime$ and A$^{\prime\prime}$ states of Rb--CH($^2\Pi$) computed from the
mutlipole expansion up to and including $R^{-10}$ and from the supermolecule
calculations is illustrated in Fig. \ref{fig2}. The agreement between the
long-range and supermolecule results is relatively good, although small 
deviations can be observed. For the A$^\prime$ state the agreement is good
at small angles, and slightly deteriorates for $\theta$ around 100$^\circ$.
The same is true for the A$^{\prime\prime}$ state, showing that our computed
coefficients are not the perfect representation of the asymptotic expansion of
the RCCSD(T) potential for this system. It should be stressed here that the long-range
coefficients reported in the present paper do not describe the asymptotics
of any potential obtained from supermolecule calculations, since for most of
the supermolecule methods the long-range asymptotics is not known. See, e.g.
Ref. \cite{Moszynski:99} for a more detailed discussion of this point. However,
the data reported in the present paper can be used in the fits of the potentials,
or in the case of lack of {\em ab initio} points at large distances, to fix
the long-range asymptotics with some switching function \cite{liesbeth}.

To illustrate the importance of the long-range coefficients with $n>6$
in Fig. \ref{fig1} we report cuts through the potential energy surfaces
of the least (Rb--CH) and most (Rb--OH) anisotropic systems for
a fixed distance $R=30$ bohr. An inspection of this figure shows that
the contribution of the coefficients beyond $n=6$ is very important.
For Rb--CH the $R^{-6}$ terms qualitatively reproduce the anisotropy of the
potential. This is not the case for Rb--OH. The inclusion of all terms
up to $n=8$ gives the correct picture of the anisotropy, and the
$R^{-9}$ and $R^{-10}$ contributions are of minor importance at this distance.
It follows from the comparison of the RCCSD(T) results with the data
computed from the asymptotic expansion, cf. Fig. \ref{fig2}, that the 
short-range exchange-repulsion effects are negligible at this distance.
Thus, our illustration of Fig. \ref{fig1} trully demonstrates the 
importance of the $R^{-8}$ and higher terms in the multipole expansion
of the interaction energy. Obviously, the importance of the contributions
beyond the $C_6$ depends on the distance $R$, but our plot clearly shows
that in the region of negligible exchange and overlap the contributions
beyond $C_6$ are important.

No literature data are available for comparison, except for the long-range
coefficients for Rb--OH obtained by Lara {\it et al.} \cite{Lara:07} by fitting
the CCSD(T) potential energy surfaces
in the A$^\prime$ and A$^{\prime\prime}$ symmetries at large distances 
to the functional form of Eq. (\ref{v2fin}). The values of the long-range
coefficients taken from Ref. \cite{Lara:07} are included in Table
\ref{tab2}. The agreement between the two sets of the results is
very reasonable. The isotropic $C_6^{00}$ coefficients agree within
7\%. The discrepancies of the anisotropic coefficients are of the
order of 10 to 15\%. Such an agreement is satisfactory given the
fact that the fitted values  effectively account for the higher
coefficients that could not be obtained from the fit. The only
significant difference is in $C_6^{22}$. Here the difference is as large as 37\%, but
this coefficient is small, and most probably could not be correctly
reproduced from the fitting procedure. By contrast, the values of
$C_7^{32}$ agree relatively well, within 10\%.

Hutson \cite{priv} estimated the lowest dispersion coefficients
$C_6^{00}$, $C_6^{20}$, and $C_6^{22}$ for Rb--OH by using the best available
data for the static polarizability of Rb and OH, and the Slater-Kirkwood
rules. He obtained an isotropic $C_6^{00}$ coefficient of 149.4 a.u.,
30\% off our value. Such an agreement is reasonable given all
the simplifications of the Slater-Kirkwood approach. The values
of the anisotropic coefficients $C_6^{20}$ and $C_6^{22}$, 4.3 and 0.9 a.u.,
respectively are four and three times  smaller than
those reported in the present paper, and thus unrealistic. 
This shows
that the applicability of simple semiempirical rules to anisotropic
interactions in open-shell complexes is of limited utility.
We have performed
a similar analysis for other complexes considered in our paper, and
came to similar conclusions.

\section{Summary and conclusions}
\label{sec5}
In the present paper we have formulated the
theory of long-range interactions between a ground-state atom in an S state
and a linear molecule in a degenerate state with a non-zero projection of the
electronic orbital angular momentum. We have shown that the
long-range coefficients describing the induction and dispersion interactions
at large atom--diatom distances
can be related to the first and second-order molecular
properties. The final expressions for the long-range coefficients were written in
terms of all components of the static and dynamic multipole polarizability
tensor, including the nonadiagonal terms connecting states with the opposite
projection of the electronic orbital angular momentum. It was also shown
that for the interactions of molecules in excited states
that are connected to the ground state by multipolar transition moments
additional terms in the long-range induction energy appears. All these
theoretical developments were illustrated with the numerical results for
systems of interest for the sympathetic cooling experiments:
interactions of the ground
state Rb($^2$S) atom with CO($^3\Pi$), OH($^2\Pi$), NH($^1\Delta$),
and CH($^2\Pi$), and of the ground state Li($^2$S) atom with CH($^2\Pi$).
Our results for the static polarizabilities of the OH and CH molecules
are in a good agreement with the {\em ab initio} data from other authors
\cite{Spelsberg:99,Pal:07}. For all systems considered in the present
paper the induction contribution to the long-range potential was found
to be important. Also the anisotropy of the long-range interaction, as
measured by the ratio $C_6^{20}/C_6^{00}$, is substantial, while the
anisotropy due to the $C_6^{22}$ is of modest importance. Relatively
good agreement between the multipole-expanded and {\em ab initio}
RCCSD(T) results was found. In the asymptotic region, where the
exchange effects are negligible, terms $R^{-n}$ with $n\ge 8$ are 
very important, and cannot be neglected. For Rb--OH we could compare
our results with the fit of {\em ab initio} RCCSD(T) points \cite{Lara:07}.
In general, relatively good agreement was found, except for the small
$C_6^{22}$ coefficient. It was also found that the Slater-Kirkwood rules
for the anisotropic long-range coefficients fail in the case of open-shell
monomers with spatial degeneracy.

\section*{Acknowledgments}
This work was supported by
the Polish Ministry of Science and Higher Education
(grant 1165/ESF/2007/03).

\newpage
\begin{table}[h]
\caption{Diagonal Cartesian components of the static dipole polarizabilities (in a.u.) for
CO($^3\Pi$), OH($^2\Pi$), NH($^1\Delta$), and CH($^2\Pi$).\label{tab0}}
\vskip 3ex
\begin{tabular}{lrcrcc}
\hline\hline
              & \clc{CO($^3\Pi$)} & \clc{OH($^2\Pi$)} & \clc{NH($^1\Delta$)} & \clc{CH($^2\Pi$)} & Reference \\
\hline
$^{{\rm x,x}}\alpha_{zz}$ & 17.97  & 8.29  &  11.23 & 15.80 & present \\
              &        & 8.75  &        &       & Ref. \cite{Spelsberg:99} \\
              &        &       &        & 15.86 & Ref. \cite{Pal:07} \\
\hline
$^{{\rm x,x}}\alpha_{xx}$ & 12.75  & 5.99  &   9.15 & 14.32 & present \\
              &        & 6.37  &        &       & Ref. \cite{Spelsberg:99} \\
\hline
$^{{\rm x,x}}\alpha_{yy}$ &  9.91  & 7.31  &   9.15 & 11.81 & present \\
              &        & 7.55  &        &       & Ref. \cite{Spelsberg:99} \\
\hline\hline
\end{tabular}
\end{table}

\newpage
           
{\small
\begin{table}[h]
\scriptsize
\caption{Long-range coefficients (in atomic units) for Rb--CO($^3\Pi$).
$C_n^{L,K}$ is the sum $C_n^{L,K}({\rm ind})+C_n^{L,K}({\rm disp})$.
The number in parentheses denotes the power of 10.\label{tab1}}
\vskip 3ex
\begin{tabular}{lrrrrrrr}
\hline\hline
$L~\rightarrow$ &\clc{0}&\clc{1}&\clc{2}&\clc{3}&\clc{4}&\clc{5}&\clc{6} \\
\hline
$C_6^{L 0}({\rm ind})$    & 1.187(2) &         & 1.187(2) &          &           &                 &\\
$C_6^{L 0}({\rm disp})$   & 3.797(2) &         & 6.400(1) &          &           &                 &\\
$C_6^{L 0}$               & 4.984(2) &         & 1.827(2) &          &           &                 &\\
\hline
$C_6^{L 2}({\rm ind})$    &          &         & 0        &          &           &                 &\\
$C_6^{L 2}({\rm disp})$   &          &         &--1.415(1)&          &           &                 &\\
$C_6^{L 2}$               &          &         &--1.415(1)&          &           &                 &\\
\hline
$C_7^{L 0}({\rm ind})$    &          & 4.646(2)&          & 3.103(2) &           &                 &\\
$C_7^{L 0}({\rm disp})$   &          & 1.045(3)&          &--1.193(2)&           &                 &\\
$C_7^{L 0}$               &          & 1.470(3)&          & 1.913(2) &           &                 &\\
\hline
$C_7^{L 2}({\rm ind})$    &          &         &          & 4.027(1) &           &                 &\\
$C_7^{L 2}({\rm disp})$   &          &         &          &--1.550(1)&           &                 &\\
$C_7^{L 2}$               &          &         &          & 2.477(1) &           &                 &\\
\hline
$C_8^{L 0}({\rm ind})$    & 7.406(3) &         & 6.165(3) &          & 1.423(3)  &                 &\\
$C_8^{L 0}({\rm disp})$   & 3.144(4) &         & 6.753(3) &          & 2.401(2)  &                 &\\
$C_8^{L 0}$               & 3.884(4) &         & 1.292(4) &          & 1.663(3)  &                 &\\
\hline
$C_8^{L 2}({\rm ind})$    &          &         & 3.168(2) &          & 1.136(2)  &                 &\\
$C_8^{L 2}({\rm disp})$   &          &         &--4.200(2)&          &--4.667(1) &                 &\\
$C_8^{L 2}$               &          &         &--1.021(2)&          &  6.695(1) &                 &\\
\hline
$C_9^{L 0}({\rm ind})$    &          & 4.166(4)&          & 1.995(4) &           &  7.102(3)       &\\
$C_9^{L 0}({\rm disp})$   &          & 1.226(5)&          &--2.757(3)&           &--2.947(3)       &\\
$C_9^{L 0}$               &          & 1.642(5)&          & 1.719(4) &           &  4.155(3)       &\\
\hline
$C_9^{L 2}({\rm ind})$    &          &         &          & 3.299(3) &           &  8.970(1)       &\\
$C_9^{L 2}({\rm disp})$   &          &         &          &--6.668(2)&           &--2.376(1)       &\\
$C_9^{L 2}$               &          &         &          & 2.632(3) &           &   6.594(1)      &\\
\hline
$C_{10}^{L 0}({\rm ind})$ &5.376(5)  &         & 4.472(5) &          & 4.844(4)  &                 & 1.609(4)\\
$C_{10}^{L 0}({\rm disp})$&9.360(5)  &         & 7.679(5) &          & 4.014(4)  &                 & --4.700(4)\\
$C_{10}^{L 0}$            &1.474(6)  &         & 1.215(6) &          & 8.858(4)  &                 & --3.091(4) \\
\hline
$C_{10}^{L 2}({\rm ind})$ &          &         & 1.592(4) &          & 2.822(4)  &                 & 3.965(3) \\
$C_{10}^{L 2}({\rm disp})$&          &         &--4.121(4)&          &--6.491(3) &                 & --1.094(3) \\
$C_{10}^{L 2}$            &          &         &  1.180(4)&          &  2.173(4) &                 & 2.862(3) \\
\hline\hline
\end{tabular}
\end{table}}

\newpage

\begin{table}[h]
\scriptsize
\caption{Long-range coefficients (in atomic units) for Rb--OH($^2\Pi$).
$C_n^{L,K}$ is the sum $C_n^{L,K}({\rm ind})+C_n^{L,K}({\rm disp})$.
The number in parentheses denotes the power of 10. The values in square brackets are the results
of Ref. \cite{Lara:07}.\label{tab2}}
\vskip 3ex
\begin{tabular}{llllrrrr}
\hline\hline
$L~\rightarrow$ &\clc{0}&\clc{1}&\clc{2}&\clc{3}&\clc{4}&\clc{5}&\clc{6} \\
\hline
$C_6^{L 0}({\rm ind})$    & 1.339(2) [1.33(2)] &         & 1.339(2) [1.33(2)] &         &           &                 &\\
$C_6^{L 0}({\rm disp})$   & 2.154(2) [1.92(2)] &         & 1.654(1) [1.80(1)] &         &           &                 &\\
$C_6^{L 0}$               & 3.494(2) [3.25(2)] &         & 1.505(2) [1.51(2)] &         &           &                 &\\
\hline
$C_6^{L 2}({\rm ind})$    &          &         &  0       &         &           &                 &\\
$C_6^{L 2}({\rm disp})$   &          &         &  3.010(0) [1.90(0)]&           &           &                 &\\
$C_6^{L 2}$               &          &         &  3.010(0) [1.90(0)]&           &           &                 &\\
\hline
$C_7^{L 0}({\rm ind})$    &          & 9.460(2)&          & 6.306(2)&           &                 &\\
$C_7^{L 0}({\rm disp})$   &          & 2.679(2)&          & 1.058(2)&           &                 &\\
$C_7^{L 0}$               &          & 1.214(3) [1.04(3)] &           & 7.365(2) [6.30(2)] &           &                 &\\
\hline
$C_7^{L 2}({\rm ind})$    &          &         &          &--4.040(1)&          &                 &\\
$C_7^{L 2}({\rm disp})$   &          &         &          &  4.800(0)&          &                 &\\
$C_7^{L 2}$               &          &         &          &--3.560(1) [--4.00(1)]&          &                 &\\
\hline
$C_8^{L 0}({\rm ind})$    & 8.518(3) &         & 8.188(3) &         & 2.117(3)  &                 &\\
$C_8^{L 0}({\rm disp})$   & 1.626(4) &         & 2.772(3) &         & 0.274(0)  &                 &\\
$C_8^{L 0}$               & 2.487(4) &         & 1.096(4) &         & 2.117(3)  &                 &\\
\hline
$C_8^{L 2}({\rm ind})$    &          &         &  3.294(2)&         &--4.442(1) &                 &\\
$C_8^{L 2}({\rm disp})$   &          &         &  3.856(2)&         &  1.397(1) &                 &\\
$C_8^{L 2}$               &          &         &  7.712(2)&         &--3.023(1) &                 &\\
\hline
$C_9^{L 0}({\rm ind})$    &          & 7.136(4)&          & 4.188(4)&           &  5.980(3)       &\\
$C_9^{L 0}({\rm disp})$   &          & 2.865(4)&          & 1.211(4)&           &  6.591(2)       &\\
$C_9^{L 0}$               &          & 1.000(5)&          & 5.399(4)&           &  6.641(3)       &\\
\hline
$C_9^{L 2}({\rm ind})$    &          &         &          &--1.811(3)&          &  1.273(2)       &\\
$C_9^{L 2}({\rm disp})$   &          &         &          &1.790(2) &           &  3.633(1)       &\\
$C_9^{L 2}$               &          &         &         &--1.632(3)&           &   1.637(2)      &\\
\hline
$C_{10}^{L 0}({\rm ind})$ &5.815(5)  &         & 5.738(5) &         & 1.457(5)  &                 &8.853(3) \\
$C_{10}^{L 0}({\rm disp})$&3.085(5)  &         & 2.046(5) &         & 2.241(4)  &                 &1.705(3)\\
$C_{10}^{L 0}$            &8.900(5)  &         & 7.784(5) &         & 1.681(5)  &                 & 1.056(4) \\
\hline
$C_{10}^{L 2}({\rm ind})$ &          &         & 2.765(4) &         &--1.320(2) &                 & 6.332(2)   \\
$C_{10}^{L 2}({\rm disp})$&          &         &  2.167(4)&         &  3.751(2) &                 & 8.513(1)  \\
$C_{10}^{L 2}$            &          &         &  4.932(4)&         &  2.431(2) &                 & 7.183(2) \\
\hline\hline
\end{tabular}
\end{table}

\newpage

\begin{table}[h]
\scriptsize
\caption{Long-range coefficients (in atomic units) for Rb--NH($^1\Delta$).
$C_n^{L,K}$ is the sum $C_n^{L,K}({\rm ind})+C_n^{L,K}({\rm disp})$.
The number in parentheses denotes the power of 10.\label{tab3}}
\vskip 3ex
\begin{tabular}{lrrrrrrr}
\hline\hline
$L~\rightarrow$ &\clc{0}&\clc{1}&\clc{2}&\clc{3}&\clc{4}&\clc{5}&\clc{6} \\
\hline
$C_6^{L 0}({\rm ind})$    & 1.120(2) &         & 1.120(2) &         &           &                 &\\
$C_6^{L 0}({\rm disp})$   & 2.849(2) &         & 2.094(1) &         &           &                 &\\
$C_6^{L 0}$               & 3.969(2) &         & 1.329(2) &         &           &                 &\\
\hline
$C_7^{L 0}({\rm ind})$    &          & 4.436(2)&          & 2.958(2)&           &                 &\\
$C_7^{L 0}({\rm disp})$   &          & 2.456(2)&          & 1.678(2)&           &                 &\\
$C_7^{L 0}$               &          & 6.892(2)&          & 4.636(2)&           &                 &\\
\hline
$C_8^{L 0}({\rm ind})$    & 6.107(3) &         & 6.830(3) &         & 1.216(3)  &                 &\\
$C_8^{L 0}({\rm disp})$   & 2.370(4) &         & 2.611(3) &         & 3.100(2)  &                 &\\
$C_8^{L 0}$               & 2.981(4) &         & 9.441(3) &         & 1.526(3)  &                 &\\
\hline
$C_8^{L 4}({\rm ind})$    &          &         &          &         & 0         &                 &\\
$C_8^{L 4}({\rm disp})$   &          &         &          &         & 1.416(3)  &                 &\\
$C_8^{L 4}$               &          &         &          &         & 1.416(3)  &                 &\\
\hline
$C_9^{L 0}({\rm ind})$    &          & 3.324(4)&          & 2.167(4)&           &  4.089(3)       &\\
$C_9^{L 0}({\rm disp})$   &          & 2.808(4)&          & 2.036(4)&           &  9.281(2)       &\\
$C_9^{L 0}$               &          & 6.132(4)&          & 4.204(4)&           &  5.017(3)       &\\
\hline
$C_9^{L 4}({\rm ind})$    &          &         &          &         &           & --7.744(2)      &\\
$C_9^{L 4}({\rm disp})$   &          &         &          &         &           &   1.442(2)      &\\
$C_9^{L 4}$               &          &         &          &         &           & --6.302(2)      &\\
\hline
$C_{10}^{L 0}({\rm ind})$ & 5.116(5) &         & 4.777(5) &         & 8.395(4)  &                 &4.889(3)\\
$C_{10}^{L 0}({\rm disp})$& 6.581(5) &         & 1.656(5) &         & 5.146(4)  &                 &1.630(3)\\
$C_{10}^{L 0}$            &1.170(6)  &         & 6.433(5) &         & 1.354(5)  &                 & 6.518(3) \\
\hline
$C_{10}^{L 4}({\rm ind})$ &          &         &          &         & 4.640(4)  &                 &1.128(3)\\
$C_{10}^{L 4}({\rm disp})$&          &         &          &         & 4.258(4)  &                 &3.692(2)\\
$C_{10}^{L 4}$            &          &         &          &         & 8.898(4)  &                 & 1.497(3) \\
\hline\hline
\end{tabular}
\end{table}

\newpage

\begin{table}[h]
\scriptsize
\caption{Long-range coefficients (in atomic units) for Rb--CH($^2\Pi$).
$C_n^{L,K}$ is the sum $C_n^{L,K}({\rm ind})+C_n^{L,K}({\rm disp})$.
The number in parentheses denotes the power of 10.\label{tab4}}
\vskip 3ex
\begin{tabular}{lrrrrrrr}
\hline\hline
$L~\rightarrow$ &\clc{0}&\clc{1}&\clc{2}&\clc{3}&\clc{4}&\clc{5}&\clc{6} \\
\hline
$C_6^{L 0}({\rm ind})$    & 9.654(1) &          &  9.654(1) &           &           &                 &\\
$C_6^{L 0}({\rm disp})$   & 3.888(2) &          &  2.844(1) &           &           &                 &\\
$C_6^{L 0}$               & 4.853(2) &          &  1.250(2) &           &           &                 &\\
\hline
$C_6^{L 2}({\rm ind})$    &          &          &  0        &           &           &                 &\\
$C_6^{L 2}({\rm disp})$   &          &          & --7.650(0)&           &           &                 &\\
$C_6^{L 2}$               &          &          & --7.650(0)&           &           &                 &\\
\hline
$C_7^{L 0}({\rm ind})$    &          & --2.974(2)&          & --1.982(2)&           &                 &\\
$C_7^{L 0}({\rm disp})$   &          &--3.874(1)&           &   3.254(2)&           &                 &\\
$C_7^{L 0}$               &          & --3.361(2) &         &   1.272(2)&           &                 &\\
\hline
$C_7^{L 2}({\rm ind})$    &          &          &           &   2.569(1)&           &                 &\\
$C_7^{L 2}({\rm disp})$   &          &          &           & --9.037(0)&           &                 &\\
$C_7^{L 2}$               &          &          &           &   1.665(1)&           &                 &\\
\hline
$C_8^{L 0}({\rm ind})$    & 6.731(3) &          &  3.341(3) &           & 8.140(2)  &                 &\\
$C_8^{L 0}({\rm disp})$   & 3.151(4) &          &  7.499(3) &           & 6.711(2)  &                 &\\
$C_8^{L 0}$               & 3.824(4) &          &  1.084(4) &           & 1.485(3)  &                 &\\
\hline
$C_8^{L 2}({\rm ind})$    &          &          &  1.170(2) &           &--7.964(0) &                 &\\
$C_8^{L 2}({\rm disp})$   &          &          & --8.110(2)&           &--5.038(1) &                 &\\
$C_8^{L 2}$               &          &          & --6.940(2)&           &--5.834(1) &                 &\\
\hline
$C_9^{L 0}({\rm ind})$    &          &--2.130(4)&          & --1.349(4)&           &  3.089(3)       &\\
$C_9^{L 0}({\rm disp})$   &          &  1.180(3)&           &   4.095(4)&           &  1.888(3)       &\\
$C_9^{L 0}$               &         & --2.012(4)&           &   2.746(4)&           &  4.977(3)       &\\
\hline
$C_9^{L 2}({\rm ind})$    &          &          &           &   1.108(3)&           & --7.006(1)      &\\
$C_9^{L 2}({\rm disp})$   &          &          &           & --9.600(2)&           & --2.901(1)      &\\
$C_9^{L 2}$               &          &          &           &   1.012(3)&           & --9.907(1)      &\\
\hline
$C_{10}^{L 0}({\rm ind})$ &3.591(5)  &          &  1.630(5) &           & 5.587(4)  &                 &3.526(3)\\
$C_{10}^{L 0}({\rm disp})$&8.679(5)  &          &  8.055(5) &           & 8.953(4)  &                 &8.714(3) \\
$C_{10}^{L 0}$            &1.227(6)  &          &  9.685(5) &           & 1.454(5)  &                 & 1.224(4) \\
\hline
$C_{10}^{L 2}({\rm ind})$ &          &          &  1.170(3) &           & 8.307(2)  &                 &7.917(2)\\
$C_{10}^{L 2}({\rm disp})$&          &          &   1.238(3)&           & 3.143(2)  &                 &--2.617(2)\\
$C_{10}^{L 2}$            &          &          &   2.408(3)&           &  1.145(3) &                 & 5.300(2) \\
\hline\hline
\end{tabular}
\end{table}

\newpage

\begin{table}[h]
\scriptsize
\caption{Long-range coefficients (in atomic units) for Li--CH($^2\Pi$).
$C_n^{L,K}$ is the sum $C_n^{L,K}({\rm ind})+C_n^{L,K}({\rm disp})$.
The number in parentheses denotes the power of 10.\label{tab5}}
\vskip 3ex
\begin{tabular}{lrrrrrrr}
\hline\hline
$L~\rightarrow$ &\clc{0}&\clc{1}&\clc{2}&\clc{3}&\clc{4}&\clc{5}&\clc{6} \\
\hline
$C_6^{L 0}({\rm ind})$    & 4.969(1) &          &  4.969(1) &           &           &                 &\\
$C_6^{L 0}({\rm disp})$   & 2.034(2) &          &  1.521(1) &           &           &                 &\\
$C_6^{L 0}$               & 2.537(2) &          &  6.491(1) &           &           &                 &\\
\hline
$C_6^{L 2}({\rm ind})$    &          &          &  0        &           &           &                 &\\
$C_6^{L 2}({\rm disp})$   &          &          & --4.227(0)&           &           &                 &\\
$C_6^{L 2}$               &          &          & --4.227(0)&           &           &                 &\\
\hline
$C_7^{L 0}({\rm ind})$    &          & --1.531(2)&          & --1.020(2)&           &                 &\\
$C_7^{L 0}({\rm disp})$   &          &--2.863(1)&           &   1.771(2)&           &                 &\\
$C_7^{L 0}$               &          & --1.817(2) &         &   7.502(2)&           &                 &\\
\hline
$C_7^{L 2}({\rm ind})$    &          &          &           &   1.322(1)&           &                 &\\
$C_7^{L 2}({\rm disp})$   &          &          &           & --5.003(0)&           &                 &\\
$C_7^{L 2}$               &          &          &           &   8.220(0)&           &                 &\\
\hline
$C_8^{L 0}({\rm ind})$    & 1.944(3) &          &  5.030(2) &           & 6.634(2)  &                 &\\
$C_8^{L 0}({\rm disp})$   & 1.125(4) &          &  3.670(3) &           & 3.547(2)  &                 &\\
$C_8^{L 0}$               & 1.319(4) &          &  4.173(3) &           & 1.018(3)  &                 &\\
\hline
$C_8^{L 2}({\rm ind})$    &          &          &  6.025(1) &           &--4.100(0) &                 &\\
$C_8^{L 2}({\rm disp})$   &          &          & --3.344(2)&           &--2.741(1) &                 &\\
$C_8^{L 2}$               &          &          & --2.742(2)&           &--3.151(1) &                 &\\
\hline
$C_9^{L 0}({\rm ind})$    &          &--4.721(3)&          & --4.619(2)&           &  1.590(3)       &\\
$C_9^{L 0}({\rm disp})$   &          &  2.573(3)&           &   1.663(4)&           &  1.019(3)       &\\
$C_9^{L 0}$               &         & --2.147(3)&           &   1.619(4)&           &  2.609(3)       &\\
\hline
$C_9^{L 2}({\rm ind})$    &          &          &           &   1.604(2)&           & --3.607(1)      &\\
$C_9^{L 2}({\rm disp})$   &          &          &           & --5.096(1)&           & --1.568(1)      &\\
$C_9^{L 2}$               &          &          &           &   1.095(2)&           & --5.175(1)      &\\
\hline
$C_{10}^{L 0}({\rm ind})$ & 6.133(4) &          &  1.983(4) &           & 1.337(4)  &                 &1.815(3)\\
$C_{10}^{L 0}({\rm disp})$& 9.074(5) &          &  3.187(5) &           & 3.728(4)  &                 &4.704(3) \\
$C_{10}^{L 0}$            & 9.687(5) &          &  3.385(5) &           & 5.065(4)  &                 & 6.519(3) \\
\hline
$C_{10}^{L 2}({\rm ind})$ &          &          &  1.424(3) &           & 5.150(2)  &                 &5.621(2)\\
$C_{10}^{L 2}({\rm disp})$&          &          &   3.582(2)&           & 2.407(2)  &                 &--1.328(2)\\
$C_{10}^{L 2}$            &          &          &   1.782(3)&           &  7.557(2) &                 & 4.293(2) \\
\hline\hline
\end{tabular}
\end{table}

\newpage
\begin{figure}
\begin{center}
\begin{minipage}{\textwidth}
\includegraphics[scale=2.70,angle=0]{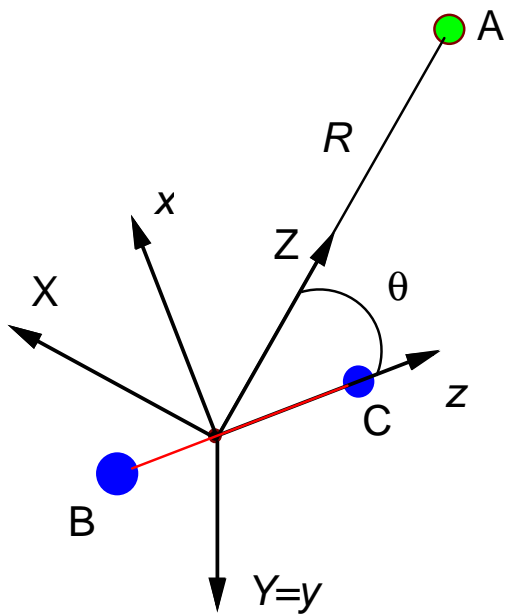}
\end{minipage}
\end{center}
\caption{
Relative orientation of the molecule-fixed frame ($xyz$) to the space-fixed frame ($XYZ$) with its $Z$ axis directed
along the vector ${\mathbf R}$ joining the center of mass of the molecule BC and the atom A. The three atoms
are in the plane $\sigma_{xz}=\sigma_{XZ}$. 
}
\label{fig3}
\end{figure}

\newpage
\begin{figure}
\begin{center}
\begin{minipage}{\textwidth}
\hspace*{-1cm}
\includegraphics[scale=0.60,angle=-90]{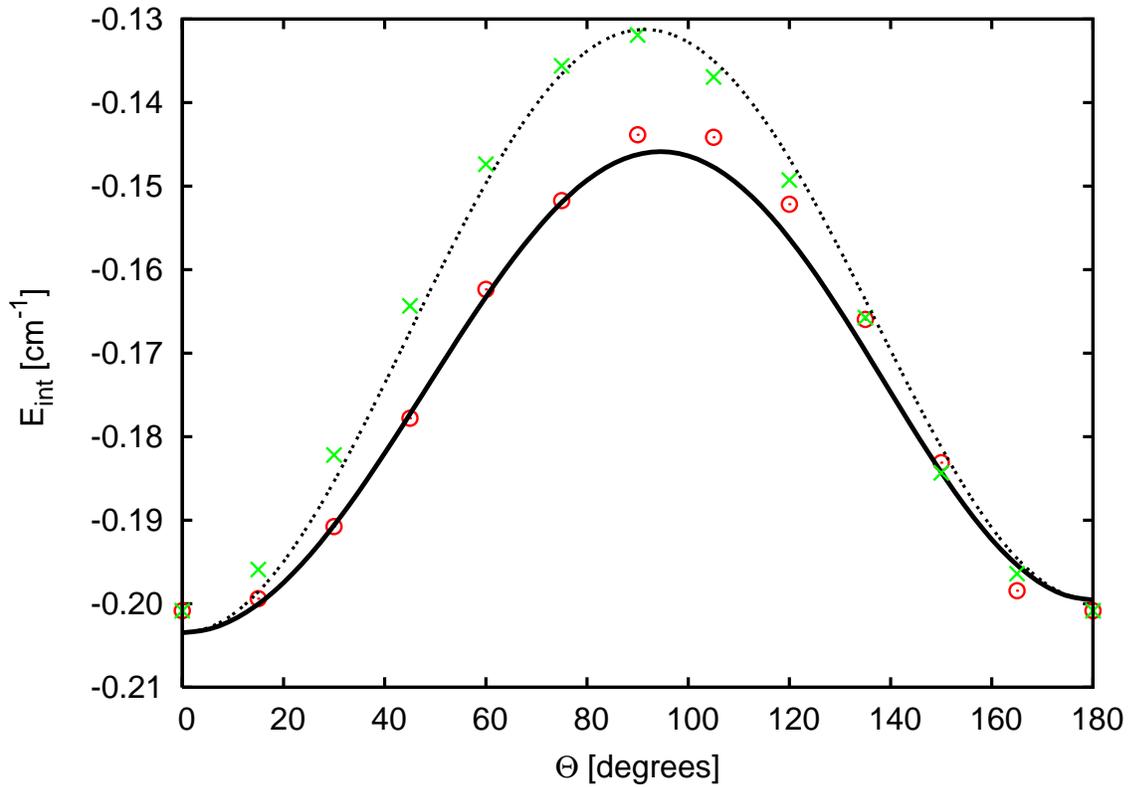}
\end{minipage}
\end{center}
\caption{Comparison of the
long-range anisotropy of the potential energy surfaces for the doublet and quartet
A$^\prime$ (solid line vs. circles) and A$^{\prime\prime}$ (dotted line vs. crosses) states of Rb--CH($^2\Pi$) computed from the
mutlipole expansion up to and including $R^{-10}$ and from the supermolecule
calculations.}
\label{fig2}
\end{figure}
\begin{figure}
\begin{center}
\begin{minipage}{\textwidth}
\hspace*{-2cm}
\includegraphics[scale=0.70,angle=-90]{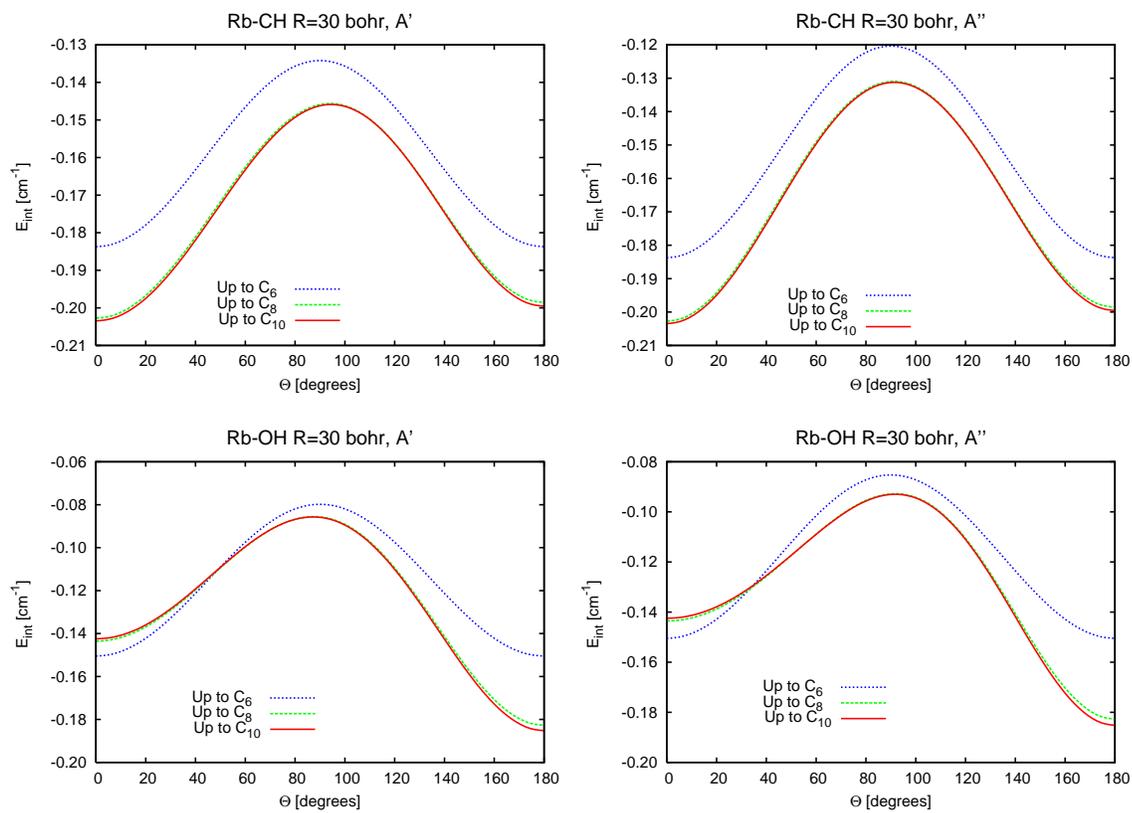}
\end{minipage}
\end{center}
\caption{Importance of the $R^{-6}$, $R^{-8}$, and $R^{-10}$ contributions to
the 
long-range anisotropy of the potential energy surfaces for the 
A$^\prime$ and A$^{\prime\prime}$ states of Rb--CH($^2\Pi$) and Rb--OH($^2\Pi$) computed from the
mutlipole expansion.}
\label{fig1}
\end{figure}
\end{document}